\documentclass[prb, twocolumn,amsmath,amssymb]{revtex4}
\usepackage{graphicx, color}
\usepackage{dcolumn}
\usepackage{bm}
\usepackage{epstopdf}
\usepackage{color}
\usepackage{subfigure}
\begin{document}
\title{Proximity Effect in Superconducting Heterostructures with
Strong Spin-Orbit Coupling and Spin Splitting} 

\author{Yao Lu}
\author{Tero T. Heikkil\"a}

\affiliation{Department of Physics and Nanoscience Center, University of Jyv\"askyl\"a,
P.O. Box 35 (YFL), FI-40014 University of Jyv\"askyl\"a, Finland}
\date{\today}
\begin{abstract}
It has been shown that singlet Cooper pairs can be converted into
triplet ones and diffuse into a ferromagnet over a long distance in a
phenomenon known as the long-range proximity effect (LRPE). This
happens in materials with inhomogeneous magnetism or spin-orbit
coupling (SOC). Most of the previous studies focus on the cases with
small SOC and exchange field. However, the physics was not clear when
SOC and exchange field strength are both much greater than the disorder strength. In this work, we consider a two dimensional system with a large Rashba-type SOC and exchange field in the case where only one band is partially occupied. We develop a generalized quasiclassical theory by projecting the Green function onto the partially occupied band (POB). We find that when the SOC energy scale is comparable with the exchange field, there is no LRPE. The reason is that the nonmagnetic impurities together with the large SOC and exchange field can effectively generate spin-flip scattering, which suppresses the proximity effect. We also show that when increasing either SOC or exchange field, the decay length of superconducting correlations can be significantly increased due to an approximately restored time reversal symmetry or spin rotation symmetry around the $z$ (out-of-plane) axis.
\end{abstract}

\maketitle
\section{INTRODUCTION}

The proximity effect in a superconductor (S)/ferromagnet (F) structure
has been extensively studied during the past decades. Experimentally a
significant increase of conductivity has been observed in the S/F
structures indicating that the Cooper pairs can penetrate into the ferromagnet
over a long distance [\onlinecite{petrashov1994conductivity,petrashov1999giant,giroud1998superconducting}]. This LRPE is unexpected because the exchange
field can destroy singlet Cooper pairs consisting of two electrons
with opposite spins. The theoretical explanation of this unusual LRPE  [\onlinecite{bergeretrmp05,PhysRevLett.86.4096,bergeretrmp18}]
is that the local inhomogeneity of magnetization in the vicinity of an S/F
interface can create triplet pairing correlations that survive in
the ferromagnet  [\onlinecite{PhysRevB.75.104509,PhysRevB.76.134502,PhysRevB.76.060504,PhysRevB.78.104509,PhysRevB.78.024519,PhysRevB.77.174511,PhysRevB.83.144515,PhysRevB.80.214537}]. The decay length of triplet pairing correlations
in the ferromagnet is of the order of the thermal coherence length $\xi_{T}=\sqrt{D/T}$ while
the singlet pairing decays over a much shorter distance $\xi_{h}=\sqrt{D/h}$ where $D$, $T$ and $h$ are diffusion constant, temperature and strength of the exchange field, respectively.
This LRPE also explains the long-range Josephson currents in S-F-S
junctions made of half metals [\onlinecite{keizer2006spin,anwar2012long}], ferromagnetic multilayers [\onlinecite{keizer2006spin}] and ferromagnets
with intrinsic inhomogeneous magnetization [\onlinecite{robinson2010controlled}]. Recently, it was realized
that the LRPE also exists in systems where the SOC operator does not commute
with the exchange field operator [\onlinecite{LRSOC1,LRSOC2}]. Interestingly, it has been shown that the system
with SOC and a uniform exchange field is actually gauge equivalent to
the system with an inhomogeneous exchange field. In particular, it has been demonstrated
that the condition for the existence of LRPE is that the effective SU(2)
electric field is finite.

The previous studies on the proximity effect in S/F hybrid structures focus
on systems with a small spin splitting field. In that case, the exchange
field and SOC can be treated perturbatively as the time and space components
of an effective SU(2) potential [\onlinecite{SU2,tokatly2017usadel,SOI}]. To take into account the effective SU(2) gauge field one just needs to replace the derivative
operator in the quasiclassical equations by a covariant derivative including the
SU(2) gauge field. The triplet pairing correlation is generated at
the interface as long as the SU(2) electric field is finite. However,
when considering systems where the exchange field and SOC strength are much
greater than the disorder strength, both exchange field and SOC have to
be treated unperturbatively, such that the conclusion drawn from the SU(2)
fields no longer applies. Whether LRPE exists in systems with a large
exchange field and SOC has remained therefore unclear.

In this work, we consider a two dimensional system with SOC and an exchange field both much
greater than the disorder strength. We assume that the spin splitting
is large enough, such that only one band is partially occupied. The
proximity effect in this system is very different from that of a system
with a nearly degenerate Fermi surface. The reason is the following.
First, due to spin-momentum locking the singlet and
triplet pairing correlations are locked together and have the same decay
length [\onlinecite{Loss2016,LRPEDirac,PETI,PETI2,RashbaSC}], in contrast to the case of a nearly degenerate Fermi surface
in which singlet and triplet pairing correlations can be treated independently.
Second, the effect of nonmagnetic impurity self-energy depends on
the spin texture on the Fermi surface, which is much more complicated
than that in the previous case. As a result, we expect a very different proximity behavior
in this large spin splitting system. In order to investigate the proximity
effect, we first develop a generalized quasiclassical theory by projecting the
Green functions onto the POB. We derive the most general normalization
condition $\hat{g}^{2}=\hat{P}$ for this quasiclassical theory where $\hat{g}$ is
the quasiclassical Green function and $\hat{P}$ is the projection operator
onto the POB. We find that there
is no LRPE when $\alpha p_{F}\approx h_{z}$, where $\alpha$ is the SOC
strength, $p_F$ is the Fermi momentum and $h_z$ is the $z$ component of the exchange field. This is because the SOC and exchange field break both spin rotation
and time reversal symmetries, such that the nonmagnetic impurities
can effectively generate spin-flip scattering, which hugely suppresses
the proximity effect. Interestingly, for an increasing SOC, the proximity
effect can be significantly enhanced until the decay length reaches
the thermal coherence length $\xi_{T}$. This is because the time reversal
symmetry is approximately restored in the limit $\alpha p_{F}\gg h_{z}$.
Similarly, the LRPE can also exist in the limit $h_{z}\gg\alpha p_{F}$
due to spin rotation symmetry around the $z$ axis.

\section{MODEL}

\begin{figure}[h!]
\centering
\subfigure{\includegraphics[width = 0.95\columnwidth]{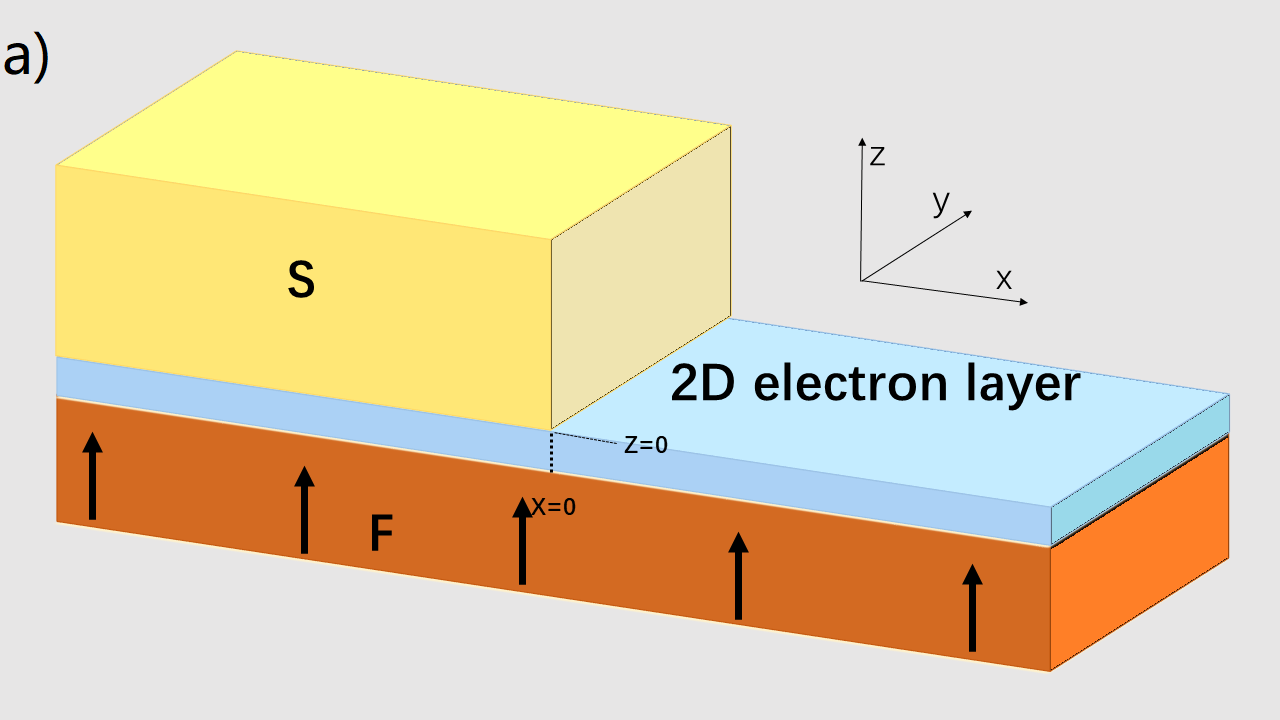}}
\subfigure{\includegraphics[width=0.95\columnwidth]{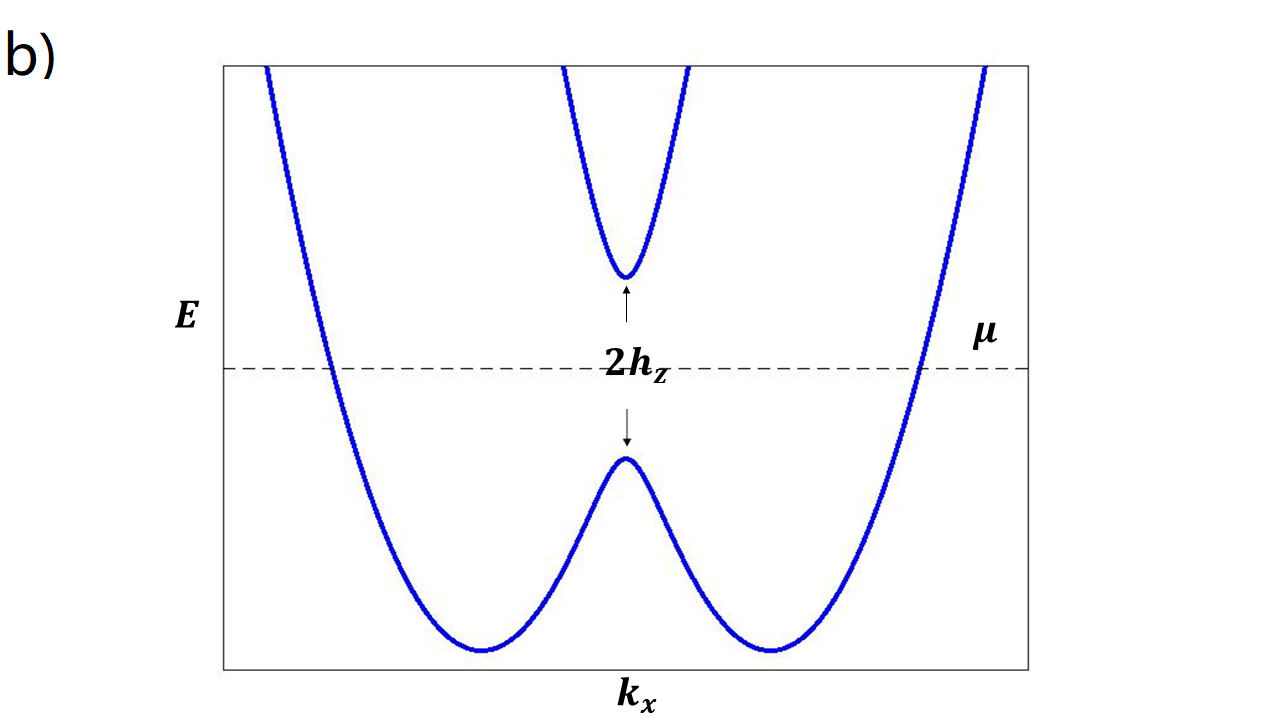}}
\caption{(a) Sketch of the system under consideration. A bulk
  superconductor induces pairing correlation in a 2D electron layer
  placed under it. (b) Schematic picture of the band structure (with
  $k_{y}=0$) of the 2D electron layer. The bands are split by the SOC
  and the exchange field, and the chemical potential is within the magnetic gap, such that the Green function can be projected to the POB.}\label{Fig:demo}
\end{figure}

We consider a 2D electron layer with a large Rashba SOC and an exchange field (induced from a ferromagnet beneath it) placed under a bulk superconductor and coupled to it via tunneling
through a thin insulator layer as shown in Fig.~\ref{Fig:demo}(a). The Hamiltonian
describing this system is given by ($\hbar=k_B=1$)

\begin{equation}
\hat{H}=\hat{H}_{S}+\hat{H}_{N}+\hat{H}_{T},
\end{equation}
where $\hat{H}_{N}$ is the Hamiltonian of a 2D electron layer

\begin{eqnarray}
\hat{H}_N=\int d^{2}r\quad c^{\dagger}(\boldsymbol{r})\Bigg[\frac{-\hat{\boldsymbol{\nabla}}^{2}}{2m_{N}}+\alpha\hat{\boldsymbol{k}}\cdot\boldsymbol{\eta}+\boldsymbol{h\cdot\sigma}\nonumber \\
+U_{N}(\boldsymbol{r})-\mu_{N}\Bigg]c(\boldsymbol{r}),
\end{eqnarray}
$\hat{H}_{S}$ describes the bulk superconductor
\begin{eqnarray}
\hat{H}_{S}=\int d^{2}rdz\Psi^{\dagger}(\boldsymbol{r},z)\bigg[\frac{-\hat{\boldsymbol{\nabla}}^{2}-\partial_{z}^{2}}{2m_{S}}+U_{S}(\boldsymbol{r})-\mu_{S}\bigg]\Psi(\boldsymbol{r},z)\nonumber\\+\Delta\hat{\Psi}_{\uparrow}^{\dagger}(\boldsymbol{r},z)\hat{\Psi}_{\downarrow}^{\dagger}(\boldsymbol{r,z})+\Delta^{*}\hat{\Psi}_{\downarrow}(\boldsymbol{r},z)\hat{\Psi}_{\uparrow}(\boldsymbol{r},z).
\end{eqnarray}
The coupling Hamiltonian $H_{T}$ has the form
\begin{equation}
\hat{H}_{T}=\int d^{2}rd^{2}r'[c^{\dagger}(\boldsymbol{r})\hat{t}\Psi(\boldsymbol{r}',0)+\Psi^{\dagger}(\boldsymbol{r}',0)\hat{t}^{\dagger}c(\boldsymbol{r})].
\end{equation}

\noindent Here $\Psi^{\dagger}(\boldsymbol{r},z)=[\Psi_{\uparrow}^{\dagger}(\boldsymbol{r},z),\Psi_{\downarrow}^{\dagger}(\boldsymbol{r},z)]$
is the creation operator of an electron in the bulk superconductor
and $c^{\dagger}(\boldsymbol{r})=[c_{\uparrow}^{\dagger}(\boldsymbol{r}),c_{\downarrow}^{\dagger}(\boldsymbol{r})]$
is the electron creation operator in the 2D electron layer. $\boldsymbol{r}$
is a 2D vector in the plane of the 2D electron layer and $z$ is the
out-of-plane coordinate. The 2D gradient is $\hat{\boldsymbol{\nabla}}=(\hat{\nabla}_x,\hat{\nabla}_y)$. $\boldsymbol{\sigma=(\sigma_{x},\sigma_{y},\sigma_{z})}$
are Pauli matrices acting on the spin space and $\boldsymbol{\eta}$ is
defined by $\boldsymbol{\eta=(-\sigma_{y},\sigma_{x})}.$ $m_{S/N}$,
$\mu_{S/N}$, $U_{S/N}$ and $\Delta$ denote the effective mass,
chemical potential, disorder potential and pairing potential in the
bulk superconductor/2D electron layer, respectively. 
$\boldsymbol{h}$ 
is the exchange field. Here we
assume that the strength of SOC and the $z$ component of the exchange field
are much greater than the disorder strength $\alpha p_{F},h_{z}\gg1/\tau$,
where $\tau$ is the scattering time, whereas the in-plane components
of $\boldsymbol{h}$ are small $h_{x},h_{y}\ll1/\tau$. We
also assume that the chemical potential is within the magnetic
gap and only one band is partially occupied as shown in Fig.~\ref{Fig:demo}(b). The S/F interface is located at $z=0$ and the right edge of the bulk superconductor is at $x=0$.
Uniform pairing correlations can be induced in the left part (with
$x<0$) and the Cooper pairs penetrate into the right part (with $x>0$)
and decay along the $x$ direction over a characteristic length $\xi$.

\begin{figure}[h!]
\centering
\includegraphics[width = 0.62\columnwidth]{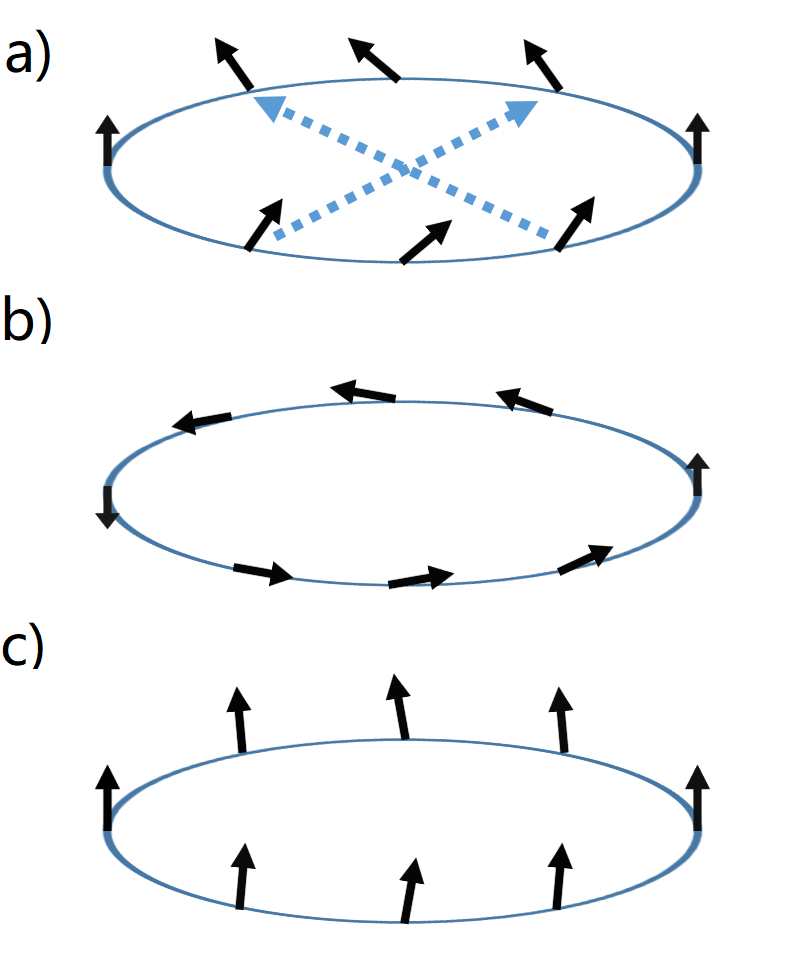}
\caption{Sketch of spin textures of the Fermi surfaces. The circles
  represent the Fermi surfaces, the solid arrows represent the
  directions of spin polarization and the dashed arrows represent the
  effective spin-flip scattering. (a) When $\alpha\approx h_z$ the
  in-plane and out-of-plane components of the spin polarization are
  comparable. (b) For $\alpha\gg h_z$ the spins are almost pinned in
  the $xy$ plane and form a helical texture. (c) When $\alpha\ll h_z$ the spins are almost polarized in the $z$ direction. }\label{Fig:Spin}
\end{figure}

In this work, our focus is to study how the pairing correlation decays
in the right part of the 2D electron layer. Before going into explicit
calculations, we can qualitively investigate the proximity effect by
analysing the properties of impurity scattering. In the case where the
SOC strength is comparable with the strength of the exchange field
$\alpha p_F\approx h_z$, the spins are on average polarized in the $z$
direction but also have considerable in-plane components as shown in
Fig.~\ref{Fig:Spin}(a). In this case, nonmagnetic impurities couple
quasi-particle states with different spins leading to an effective
spin-flip scattering [dashed arrows in Fig.~\ref{Fig:Spin}(a)], which
suppresses the proximity effect. In this case LRPE does not take
place. In the large SOC limit $\alpha\gg h_z$, the spins are almost
pinned in the $xy$ plane and form a helical texture
[Fig.~\ref{Fig:Spin}(b)]. The electrons with opposite momenta also
have nearly opposite spins, which means that the time reversal
symmetry is approximately preserved. Therefore, the electron
scattering that does not break the time reversal symmetry cannot lead
to a large effective spin-flip scattering, allowing for the presence
of LRPE. In the large exchange field limit $h_z\gg\alpha p_F$, the spins are almost polarized in the $z$ direction and no spin can be flipped [\onlinecite{Mihail}] [Fig.~\ref{Fig:Spin}(c)]. Thus LRPE is allowed if equal spin pairing is formed.

\section{QUASICLASSICAL THEORY}
In order to quantitively study the proximity effect, we first develop
a quasiclassical theory for the 2D electron layer and then include
the effect of the bulk superconductor as a boundary condition. We start
with the Gorkov equation for the right part of the 2D electron layer.
Within Born approximation and in spin$\otimes$particle-hole space, the Gorkov
equation can be written as
\begin{equation}
(\hat{G}_{0}^{-1}+\mu-\hat{\Sigma})\hat{G}=1 \label{Eq:Gorkov}
\end{equation}
\begin{equation}
\hat{G}_{0}^{-1}=-\frac{\hat{\boldsymbol{k}}^{2}}{2m_{N}}-\alpha\hat{\boldsymbol{k}}\cdot\boldsymbol{\eta}+(i\omega_{n}+\boldsymbol{h}\cdot\boldsymbol{\sigma})\tau_{3}.
\end{equation}

Here, $\omega_{n}$ is the Matsubara frequency $\omega_{n}=(2n+1)\pi T$
with $n=0,1,2\cdots$, $\tau_{3}$ is the Pauli matrix acting on the particle-hole basis and $\hat{\Sigma}$ is the disorder self-energy obtained
from Born approximation, $\text{\ensuremath{\hat{\Sigma}=\langle \hat{G}\rangle/\tau}}$, where $\langle\cdot\rangle$ means an angular average over the momentum directions. In order to solve the Gorkov equation, we perform the quasiclassical approximation
and obtain the Eilenberger equation [\onlinecite{eilenberger1968transformation,1999quasiclassical}]

\begin{eqnarray}
\frac{p_{F}}{m_{N}}\boldsymbol{\nabla}\hat{g}+\frac{\alpha}{2}\{\boldsymbol{\eta},\boldsymbol{\nabla}\hat{g}\}=\Bigg[\hat{g},\omega_{n}\tau_{3}+i\boldsymbol{h}'\cdot\boldsymbol{\sigma'}+ih_{z}\sigma_{3}\nonumber\\+i\alpha p_{F}\boldsymbol{\eta}\cdot\boldsymbol{n}_{F}+\frac{\langle \hat{g}\rangle}{2\tau}\Bigg].\label{Eq:Eilenberger}
\end{eqnarray}
Here $\boldsymbol{h}'$ and $\boldsymbol{\sigma}'$ are in-plane components of the exchange field and Pauli matrices, respectively, defined by $\boldsymbol{h}'=(h_x,h_y)$, $\boldsymbol{\sigma}'=(\sigma_x,\sigma_y)$. We have performed a Fourier transformation of the Green function
in Eq.~\eqref{Eq:Gorkov} with respect to the relative space argument and then taken
the integral over $\varepsilon_{p}=p^{2}/2m_{N}-S-\mu_{N}$ with $S=\sqrt{h_{z}^{2}+\alpha^{2}p^{2}}$,
which results in 
\begin{equation}
\hat{g}(\omega_{n};\boldsymbol{R},\boldsymbol{n}_{F})=\int\frac{d\varepsilon_{p}}{i\pi}\hat{G}(\omega_{n};\boldsymbol{R},\boldsymbol{p}).\label{Eq:quasiclassical}
\end{equation}
Here $\text{\ensuremath{\boldsymbol{n}_{F}=\boldsymbol{p}_{F}/|\boldsymbol{p}_{F}|}}$is
a unit vector in the direction of momentum $\boldsymbol{p}_{F}$ at the Fermi level. Denoting $\hat{Q}\equiv ih_{z}\sigma_{3}+i\alpha p_{F}\boldsymbol{\eta}\cdot\boldsymbol{n}_{F}$,
we emphasize that the main difference between Eq.~\eqref{Eq:Eilenberger}
and an ordinary Eilenberger equation is that on the right hand side the
dominant term is $\hat{Q}$ rather than $\langle \hat{g}\rangle/\tau$.

Since the dominant term on the right hand side of Eq.~\eqref{Eq:Eilenberger} is $\hat{Q}$,
the leading term of the quasiclassical Green function $\hat{g}$ should
commute with $\hat{Q}$ and all other terms are of the order of $1/\tau S_F\ll1$ where $S_F$ is the absolute value of the eigenvalue of $\hat{Q}$ at the Fermi level, $S_F=\sqrt{\alpha^2p_F^2+h_z^2}$. Using the condition of $[\hat{g},\hat{Q}]=0$, we can write
down the most general form of $\hat{g}$
\begin{eqnarray}
\hat{g}=\sum_{\lambda=\pm}a_{\lambda}|\psi_{\text{\ensuremath{\lambda,}}e}\rangle\langle\psi_{\lambda,e}|+b_{\lambda}|\psi_{\lambda,h}\rangle\langle\psi_{\text{\ensuremath{\lambda,h}}}|\nonumber\\+d_{\lambda}|\psi_{\lambda,e}\rangle\langle\psi_{\lambda,h}|+e_{\lambda}|\psi_{\lambda,h}\rangle\langle\psi_{\lambda,e}|,
\end{eqnarray}
where $|\psi_{\lambda,e/h}\rangle$ are eigenvectors of operator $\hat{Q}$
satisfying $\hat{Q}|\psi_{\lambda,e/h}\rangle=\lambda iE|\psi_{\lambda,e/h}\rangle$
with $\lambda=\pm$ being band indices and $e/h$ labeling particle
and hole eigenvectors. As the upper band is very far away from the
chemical potential, the quasiclassical Green function $\hat{g}$ should
contain no information of the upper band. Therefore, we can safely
drop the $|\psi_{+,e/h}\rangle$ terms in $\hat{g}$ and write it as 
\begin{eqnarray}
\hat{g}=a|\psi_{\text{\ensuremath{-,}}e}\rangle\langle\psi_{-,e}|+b|\psi_{-,h}\rangle\langle\psi_{\text{\ensuremath{-,h}}}|\nonumber\\+d|\psi_{-,e}\rangle\langle\psi_{-,h}|+e|\psi_{-,h}\rangle\langle\psi_{-,e}|
\end{eqnarray}
with
\begin{equation}
|\psi_{-,e}\rangle=\begin{pmatrix}\begin{array}{cccc}
-\alpha p_{F}e^{i\phi/2}, & (S_{F}+h_{z})e^{-i\phi/2}, & 0, & 0\end{array}\end{pmatrix}^{\text{T}}/N
\end{equation}
\begin{equation}
|\psi_{-,h}\rangle=\begin{pmatrix}\begin{array}{cccc}
0, & 0, & (S_{F}+h_{z})e^{i\phi/2}, & -\alpha p_{F}e^{-i\phi/2}\end{array}\end{pmatrix}^{\text{T}}/N,
\end{equation}
where $a,b,d,e$ are arbitary constants, $N$ is the normalization
factor $N=\sqrt{2S_{F}(S_{F}+h_{z})}$ and
$\phi$ is the angle between $\boldsymbol{n}_{F}$ and the $y$ axis so that $\cos(\phi)=n_{y},\sin(\phi)=n_{x}$. By doing this we actually
project the Green function onto the POB. Next we write Eq.~\eqref{Eq:Eilenberger} in
the new basis $\psi_{-}=(\psi_{-,e},\psi_{-,h})^{\text{T}}$ and obtain
the effective low energy Eilenberger equation
\begin{equation}
v_{F}\boldsymbol{\nabla}\hat{g}=[\hat{g},w_n\tilde{\tau}_{3}+i\frac{\alpha p_{F}}{S_{F}}(\boldsymbol{h}'\times\hat{z})\cdot\boldsymbol{n}_{F}+\langle \hat{g}\rangle/2\tau]. \label{Eq:EilenbergerPOB}
\end{equation}

Here $\hat{z}$ is the unit vector in the $z$ direction. Equation \eqref{Eq:EilenbergerPOB} cannot uniquely determine the quasiclassical Green function
$\hat{g}$ and has to be supplemented by a normalization condition. However,
we cannot simply borrow the ordinary normalization condition $\hat{g}^{2}=\hat{1}$
as $\hat{g}$ lives only in a subspace of the whole Hilbert space. In order
to find the normalization condition for this model, we project the
single particle bulk Green function onto the POB and apply the quasiclassical
approximation, which leads to 
\begin{equation}
\hat{G}=\frac{\hat{P}}{\hat{P}(i\omega_{n}\tau_{3}+iv_{F}\partial_{r}-U)\hat{P}-\varepsilon_{p}\hat{P}}
\end{equation}
where $\hat{P}$ is given by $\hat{P}=|\psi_{-,e}\rangle\langle\psi_{-,e}|+|\psi_{-,h}\rangle\langle\psi_{-,h}|$.
We define $\hat{O}=\hat{P}(i\omega_{n}\tau_{3}+iv_{F}\partial_r-U)\hat{P}$ and have
$\hat{G}=1/(\hat{O}-\varepsilon_{p})$. According to Eq.~\eqref{Eq:quasiclassical}, the principal value
integral along the real $\varepsilon_{p}$ axis is equal to the sum
of two integrals along contours which consist of the real axis (from
$-\infty$ to $+\infty$) closed by semicircles in the upper and lower
half-planes $\int=\frac{1}{2}\oint_{up}+\frac{1}{2}\oint_{dn}$. Working
out the contour integral, we obtain [\onlinecite{eckern1981quasiclassical}]
\begin{equation}
\hat{g}=\hat{P}_{up}-\hat{P}_{dn}.
\end{equation}

Here, $\hat{P}_{up/dn}$ is the projection operator defined by $\hat{P}_{up/dn}=|\psi_{up/dn}\rangle\langle\psi_{up/dn}|$
and $|\psi_{up/dn}\rangle$ is the eigenvector of operator $\hat{O}$ with
eigenvalue having the positive/negative imaginary part. Therefore, we
have $\hat{g}^{2}=\hat{P}_{up}+\hat{P}_{dn}$. Since $\hat{P}_{up}+\hat{P}_{dn}$ is the identity
operator in the $\psi_{-}$ subspace, we immediately have $\hat{P}_{up}+\hat{P}_{dn}=\hat{P}$.
Thus the normalization condition reads 
\begin{equation}
\hat{g}^{2}=\hat{P}.
\end{equation}

For convenience, we can write $\hat{g}$ in a more symmetric form $\hat{g}=\sum_{i=0}^{3}g_{i}\tilde{\tau_{i}}$
with $\tilde{\tau}_{0}=\hat{P}$ and other $\tilde{\tau}_{i}$ being Pauli
matrices acting on the $\psi_{-}$ basis. In terms of the coefficients
$g_{i}$ , the normalization condition can be written as $g_{0}=0$
, $g_{1}^{2}+g_{2}^{2}+g_{3}^{2}=1$.

\section{PROXIMITY EFFECT IN the DIFFUSIVE LIMIT}

Here, we consider the diffusive limit $1/\tau\gg\omega,h_{x},h_{y},\Delta.$
In this limit, the dominant term on the right hand side of Eq.~\eqref{Eq:EilenbergerPOB}
is $\langle \hat{g}\rangle/2\tau$. Thus the leading term $\hat{g}_{0}$ of the quasiclassical
Green function $\hat{g}$ satisfies
\begin{equation}
[\hat{g}_{0},\langle \hat{g}_{0}\rangle/2\tau]=0.
\label{eq:selfconsistency}
\end{equation}
We need to be careful when calculating $\langle \hat{g}_{0}\rangle$ since
the basis $\tilde{\tau}$ is now $\boldsymbol{n}_{F}$ dependent.
The proper way to do the angular average is to write $\tilde{\tau}$
in the usual spin$\otimes$particle-hole basis, perform the angular average for the
4 $\times$ 4 matrix and then project it back onto the $\psi_{-}$ space. We
first consider the case $\alpha p_{F}\approx h_{z}$. In this case,
we find that the only solution of Eq.~\eqref{eq:selfconsistency} is $\hat{g}_{0}=g_{s,3}\tilde{\tau}_{3}$
where $g_{s,3}$ is $\boldsymbol{n}_{F}$ independent constant (details can be found in Appendix. A). The
$\tilde{\tau}_{1}$ and $\tilde{\tau}_{2}$ terms which represent
pairing correlations do not appear in the leading term $\hat{g}_{0}$. This
means that there is almost no proximity effect when $\alpha p_{F}$
is comparable with $h_{z}$. The physical picture is that in this
case, the time reversal symmetry and spin rotation symmetry are both
broken and the disorder potential can effectively generate spin-flip scattering, which destroys the LRPE. In order to find the solution
of $\hat{g}_{0}$ with a finite off-diagonal term, we need to consider
the limits $\alpha p_{F}\gg h_{z}$ and $\alpha p_{F}\ll h_{z}$.
In the former case, the system has an approximate time reversal symmetry
and in the latter case the spin rotation symmetry around the $z$ axis is approximately
restored.

\subsection{Large SOC Limit}

Considering the limit $\alpha p_{F}\gg h_{z}$, we can find one approximate
solution of Eq.~\eqref{eq:selfconsistency}, which is $\hat{g}_{0}=\hat{g}_{s}=\sum_{i}g_{s,i}\tilde{\tau}_{i}$
where $g_{s,i}$ are $\boldsymbol{n}_{F}$ independent coefficients.
The angular average is calculated as $\langle \hat{g}_{0}\rangle=\frac{1}{2}\hat{g}_{0}+\frac{X}{2}\tilde{\tau}_{3}\hat{g}_{0}\tilde{\tau}_{3}$
with $X=h_{z}^{2}/S_{F}^{2}$ (see Appendix. B). Hence we have 

\begin{equation}
[\hat{g}_{0},\langle \hat{g}_{0}\rangle/2\tau]=\frac{X}{2}[\hat{g}_{0},\tilde{\tau}_{3}\hat{g}_{0}\tilde{\tau}_{3}/2\tau].    
\end{equation}

It can be seen that this solution does not strictly satisfy Eq.~\eqref{eq:selfconsistency}
but only with a small error of the order of $X$. This error is acceptable
since we already drop some small terms of the order of $\omega_{n}\tau$
in deriving Eq.~\eqref{eq:selfconsistency}. Then the Green function can be expanded up to
the first two terms of the 2D harmonics $\hat{g}=\hat{g}_{s}+\boldsymbol{n}_{F}\cdot\hat{\boldsymbol{g}}_{a,i}=\sum_{i}(g_{s,i}+\boldsymbol{n}_{F}\cdot\boldsymbol{g}_{a,i})\tilde{\tau}_{i}$
where the zeroth harmonic is isotropic and its amplitude is much larger
than that of the first harmonic. Substituting this expansion into Eq.~\eqref{Eq:EilenbergerPOB}
and taking an average over all directions of $\boldsymbol{n}_{F}$
in the coefficients, we obtain
\begin{equation}
\frac{1}{2}v_{F}\tilde{\boldsymbol{\nabla}}\hat{\boldsymbol{g}}_{a}=[\hat{g}_{s},\omega_{n}\tilde{\tau}_{3}+\frac{X}{2}\tilde{\tau}_{3}\hat{g}_{s}\tilde{\tau}_{3}/2\tau], \label{Eq:average1}
\end{equation}
where $\tilde{\boldsymbol{\nabla}}$ is the covariant derivative $\tilde{\boldsymbol{\nabla}}\boldsymbol{\cdot}=\boldsymbol{\nabla}\boldsymbol{\cdot}+\frac{i\alpha p_{F}}{S_{F}v_{F}}\hat{z}\times\boldsymbol{h'}[\tilde{\tau}_{3},\boldsymbol{\cdot}]$,
in which the small in-plane components of the exchange field $h_{x}$
and $h_{y}$ play the role of the $U(1)$ gauge field. Multiplying Eq.~\eqref{Eq:EilenbergerPOB}
by $\boldsymbol{n}_{F}$ and taking the angular average for the coefficients,
we obtain 
\begin{equation}
\tau v_{F}\tilde{\nabla}\hat{g}_{s}=\frac{1}{4}\hat{\boldsymbol{g}}_{a}\hat{g}_{s},\label{Eq:average2}
\end{equation}
where we have used the fact that $X\ll1$ , $w_{n}\ll1/\tau$. Combining
Eq.~\eqref{Eq:average1} and~\eqref{Eq:average2}, we arrive at the Usadel equation [\onlinecite{usadel1970generalized}]
\begin{equation}
4D\tilde{\boldsymbol{\nabla}}(\hat{g}_{s}\tilde{\boldsymbol{\nabla}}\hat{g}_{s})=[\omega_{n}\tilde{\tau}_{3}+\frac{X}{4}\tilde{\tau}_{3}\hat{g}_{s}\tilde{\tau}_{3}/\tau,\hat{g}_{s}], \label{Eq:Usadel1}
\end{equation}
\noindent where the diffusion constant $D$ is given by $D=\tau v_F^2/2$. Compared to the usual case, we have an extra term $\frac{X}{4}\tilde{\tau}_{3}\hat{g}_{s}\tilde{\tau}_{3}/\tau$,
which represents an effective spin-flip scattering due to impurities coupling different spin-momentum locked states [\onlinecite{RevModPhys1986}]. It is
a small term proportional to $X$ and does not totally destroy the
proximity effect. We consider the weak proximity limit and write the Green
function as $\hat{g}_{s}\approx\tilde{\tau}_{3}+\hat{f}=\tilde{\tau}_{3}+f_{1}\tilde{\tau}_{1}+f_{2}\tilde{\tau}_{2}$.
Substituting this expansion into Eq.~\eqref{Eq:Usadel1}, we have
\begin{equation}
4\tau D\tilde{\nabla}^{2}\hat{f}=X\hat{f}+2\omega_{n}\tau \hat{f}.
\end{equation}
Solving this equation, we get 
\begin{equation}
\hat{f}=f_{0}\begin{pmatrix}\begin{array}{cc}
0 & e^{-2i\alpha h_{y}x/S_{F}v_{F}}\\
e^{2i\alpha h_{y}x/S_{F}v_{F}} & 0
\end{array}\end{pmatrix}e^{-\kappa x}\quad(x>0),\label{Eq:solution1}
\end{equation} 
where $f_{0}$ is a constant determined by the boundary condition and
\begin{equation}
\kappa=\sqrt{\frac{X}{4\tau D}+\frac{\omega_{n}}{2D}+\frac{4h_{x}^{2}\alpha^{2}p_{F}^{2}}{S_{F}^{2}v_{F}^{2}}}.\label{Eq:kappa1}
\end{equation}

The Pauli matrices $\tilde{\tau}_{1/2}$ written in the usual spin$\otimes$particle-hole space are
given by

\begin{align}
&\tilde{\tau}_{+}=\tilde{\tau}_{-}^{\dagger}=(\tilde{\tau}_{1}+i\tilde{\tau}_{2})/2\nonumber\\&\hat{=}\begin{pmatrix}\begin{array}{cccc}
0 & 0 & -\alpha p_{F} & (S_{F}-h_{z})e^{i\phi}\\
0 & 0 & (S_{F}+h_{z})e^{-i\phi} & -\alpha p_{F}\\
0 & 0 & 0 & 0\\
0 & 0 & 0 & 0
\end{array}\end{pmatrix}/2S_{F}.
\end{align}

The $\alpha p_{F}$ terms represent the singlet pairing correlations while the $S_{F}+h_{z}$ and $S_{F}-h_{z}$ terms represent the triplet pairing.
One can see that the induced pairing correlation is a combination
of singlet and triplet pairing, which are locked together
and have the same decay length $\xi=1/\kappa\approx\min\left\{\sqrt{\frac{4\tau D}{X}},\sqrt{\frac{2D}{\omega_{n}}},\frac{S_{F}v_{F}}{2h_{x}\alpha p_{F}}\right\}$.
It can be seen that the different components of the exchange field play
different roles in the proximity effect. $h_{z}$ and $h_{x}$ tend
to suppress the decay length to be $\sqrt{\frac{4\tau D}{X}}$ and
$\frac{S_{F}v_{F}}{2h_{x}\alpha p_{F}}$, respectively, while $h_{y}$ does not affect
the decay length but introduces a phase gradient in the induced
pairing correlations, similar to the anomalous Josephson effect induced by the SOC and Zeeman effect [\onlinecite{AJ1,AJ2,AJ3,AJ4,AJ5,AJ6}]. In the case of no $h_{x}$ and relatively small
$X$, the decay length becomes $\xi\approx$ $\sqrt{\frac{4\tau D}{X}}$.
The decay length thus depends directly on the time reversal symmetry breaking
factor $X$. When decreasing $X$, the time reversal symmetry is further
restored and the proximity effect is promoted until the decay length
reaches the thermal coherence length $\xi_{T}$. When increasing $X$,
the time reversal symmetry is further broken, and the decay length
is suppressed to the mean free path $l=\tau v_{F}$. Thus the system
can smoothly cross over to the $\alpha p_{F}\approx h_{z}$ regime, where the decay length is of the order of $l$.

\subsection{Large Exchange Field Limit}

Next, we consider the opposite limit $h_{z}\gg\alpha p_{F}$. In this
limit, the spins of the Fermi surface are almost polarized in the $z$  direction [Fig.~\ref{Fig:Spin}(c)]. Thus
the spin rotation symmetry around the $z$ axis is almost restored. In this
case, we can find another approximate solution of Eq.~\eqref{eq:selfconsistency}, which
is 
\begin{equation}
\hat{g}_{0}=g_{s,+}e^{i\phi}\tilde{\tau}_{+}+g_{s,-}e^{-i\phi}\tilde{\tau}_{-}+g_{s,3}\tilde{\tau}_{3},
\end{equation}
where $\tilde{\tau}_{+}$ and $\tilde{\tau}_{-}$ are defined by $\tilde{\tau}_{+}=(\tau_{1}+i\tau_{2})/2$,
$\tilde{\tau}_{-}=(\tilde{\tau}_{1}-i\tilde{\tau}_{2})/2$, and $g_{s,+}$
$g_{s,-}$ $g_{s,3}$ are $\boldsymbol{n}_{F}$ independent constants.
Unlike the large SOC limit, we have here a phase factor in the $\tilde{\tau}_{-}$
and $\tilde{\tau}_{+}$ terms. It is convenient to absorb the phase
factor into the basis by defining a new basis $|\psi '_{-}\rangle=(|\psi '_{-,e}\rangle,|\psi '_{-,h}\rangle)^{\text{T}}=(e^{i\phi/2}|\psi_{-,e}\rangle,e^{-i\phi/2}|\psi_{-,h}\rangle)^{\text{T}}.$
Then the solution can be written in this new basis as
\begin{equation}
\hat{g}_{0}=\sum_{i}g_{s,i}\tilde{\tau}'_{i},
\end{equation}
where $\tilde{\tau}'_{i}$ are Pauli matrices acting on the $|\psi '_{-}\rangle$
basis and $g_{s,i}$ are constants. The angular average of $\hat{g}_{0}$
becomes $\langle \hat{g}_{0}\rangle=\hat{g}_{0}+\frac{Y}{2}\tilde{\tau}'_{3}\hat{g}_{0}\tilde{\tau}'_{3}$
with $Y=(S_{F}-h_{z})^{2}/4S_{F}^{2}$ (see Appendix. B). Using the same method as for the large SOC limit, we get the Usadel equation

\begin{equation}
D\tilde{\boldsymbol{\nabla}}(\hat{g}_{s}\tilde{\boldsymbol{\nabla}}\hat{g}_{s})=[\omega_{n}\tilde{\tau}'_{3}+\frac{Y}{4}\tilde{\tau}'_{3}\hat{g}_{s}\tilde{\tau}'_{3}/\tau,\hat{g}_{s}].
\end{equation}

Again in the weak proximity limit, we obtain the pair correlation in the
magnet as

\begin{equation}
\hat{f}=f_{0}\begin{pmatrix}\begin{array}{cc}
0 & e^{-2i\alpha h_{y}x/S_{F}v_{F}}\\
e^{2i\alpha h_{y}x/S_{F}v_{F}} & 0
\end{array}\end{pmatrix}e^{-\kappa x}\quad(x>0)\label{Eq:solution2}
\end{equation}
with 
\begin{equation}
\kappa=\sqrt{\frac{Y}{\tau D}+\frac{2\omega_{n}}{D}+\frac{4h_{x}^{2}\alpha^{2}p_{F}^{2}}{S_{F}^{2}v_{F}^{2}}}.
\end{equation}

The Pauli matrices $\tilde{\boldsymbol{\tau}}'$ written in the usual spin$\otimes$particle-hole space are
given by
\begin{align}
&\tilde{\tau}'_{+}=\tilde{\tau}_{-}'^{\dagger}=(\tilde{\tau}'_{1}+i\tilde{\tau}'_{2})/2\nonumber\\&\hat{=}\begin{pmatrix}\begin{array}{cccc}
0 & 0 & -\alpha p_{F}e^{i\phi} & (S_{F}-h_{z})e^{i2\phi}\\
0 & 0 & (S_{F}+h_{z}) & -\alpha p_{F}e^{i\phi}\\
0 & 0 & 0 & 0\\
0 & 0 & 0 & 0
\end{array}\end{pmatrix}/2S_{F}.
\end{align}

Therefore, there exists LRPE with a decay length $\xi=1/\kappa\approx\min\left\{\sqrt{\frac{\tau D}{Y}},\sqrt{\frac{D}{2\omega_{n}}},\frac{S_{F}v_{F}}{2h_{x}\alpha p_{F}}\right\}$ dominated by triplet pairing. Again in the case of no $h_{x}$
and relatively small $Y$, the decay length becomes $\xi\approx$
$\sqrt{\frac{\tau D}{Y}}$. It can be seen that now the spin rotation
symmetry breaking factor $Y$ takes the similar role of the time reversal
symmetry breaking factor $X$ in the opposite limit. Without $h_{x}$, the decay length
can be increased up to the thermal coherence length when $Y$ is decreased
due to the further restored spin rotation symmetry around the $z$ axis. On the other
hand, when increasing $Y$, the decay length is suppressed
to the mean free path $l=\tau v_{F}$ and the system again can smoothly
cross over to the $\alpha p_{F}\approx h_{z}$ regime.

\section{BOUNDARY CONDITION}

In this section, we show how to calculate the constant $f_{0}$ in
Eq.~\eqref{Eq:solution1} and~\eqref{Eq:solution2} through a boundary condition. In this system the superconducting proximity effect contains two processes:
one is inducing uniform pairing correlations in the left part ($x<0$)
of the 2D electron layer and the other is the Cooper pairs penetrating
into the right part ($x>0$). To calculate the induced pairing correlation
in the left part, we add the self-energy of the bulk superconductor
into the Gorkov equation of the 2D electron layer following Refs.~[\onlinecite{BC1,BC2,PETI}].
The Gorkov equation for the left part of the 2D electron layer reads
\begin{equation}
(\hat{G}_{0}^{-1}+\mu-\hat{\Sigma}-\hat{\Sigma}_{S})\hat{G}=1.\label{Eq: Gorkovleft}
\end{equation}
Here, $\hat{G}_{0}^{-1}$, $\mu$ and $\Sigma$ are the same as defined
in Sec. II. $\Sigma_{S}$ is the self-energy of the bulk superconductor
given by
\begin{equation}
\hat{\Sigma}_{S}=\hat{t}^{\dagger}(\boldsymbol{p})\hat{G}_{S}(\boldsymbol{p},z=0)\hat{t}(\boldsymbol{p}),\label{Eq: selfenergyS}
\end{equation}
where $\boldsymbol{p}$ is a 2D vector in the plane of the 2D electron
layer. We assume translational invariance for the bulk superconductor
and the left part of the electron layer, such that the Green function
$\hat{G}_{S}$ has only one momentum parameter. $\hat{G}_{S}(\boldsymbol{p},z=0)$
is calculated as
\begin{equation}
\hat{G}_{S}(\boldsymbol{p},z=0)=\int\frac{dp_{z}}{2\pi}\hat{G}_{S}(\boldsymbol{p},p_{z})
\end{equation}
and $\hat{G}_{S}(\boldsymbol{p},p_{z})$ is determined by the Born self-consistency
equation
\begin{equation}
\left(i\omega_{n}\tau_{3}-\frac{\boldsymbol{p}^{2}+p_{z}^{2}}{2m_{S}}-\Delta\tau_{1}+\mu_{S}-\frac{i}{2\tau_{S}}\langle \hat{G}_{S}\rangle\right)\hat{G}_{S}=1, \label{Eq: Born}
\end{equation}
where $\text{\ensuremath{\tau_{S}}}$ is the impurity scattering time
in the bulk superconductor. The tunneling operator $\hat{t}$ is in
general a 4 $\times$ 4 matrix in spin$\otimes$particle-hole space. In the limit
$t/S_F\ll1$, only the quasiparticles
in the $\psi_{-}$ subspace can tunnel into the 2D electron layer, which
means that the tunneling matrix has the form
\begin{equation}
\hat{t}=t\hat{P}_t,\label{Eq:hopping}
\end{equation}
where $t$ is the tunneling amplitude. $\hat{P}_t$ is defined by $\hat{P}_t=|\psi_{-,e}\rangle\langle\psi_{-,e}|+|\psi_{-,h}\rangle\langle\psi_{-,h}|$ for $\alpha p_F\gg h_z$ and $\hat{P}_t=|\psi_{-,e}\rangle\langle\psi '_{-,e}|+|\psi_{-,h}\rangle\langle\psi '_{-,h}|$ for $\alpha p_F\ll h_z$. Substituting Eq.~(\ref{Eq: selfenergyS}-\ref{Eq:hopping}) into Eq.~\eqref{Eq: Gorkovleft}, performing
the quasiclassical approximation and projecting onto the POB, we obtain the
Eilenberger equation for the left part of the 2D electron layer
\begin{equation}
v_{F}\nabla \hat{g}=[\hat{g},w\tilde{\tau}_{3}+i\frac{\alpha p_{F}}{S_{F}}(\boldsymbol{h}'\times\hat{z})\cdot\boldsymbol{n}_{F}+t^2\Delta'\tilde{\tau}_{1}+\langle \hat{g}\rangle/2\tau]\quad(x<0)\label{Eq:EilenbergerLeft}
\end{equation}
Here, we have used the tunneling
condition $t/\mu_{N}\ll 1$, such that the renormalization of $\mu_{N}$
by the bulk superconductor is negligible and the only effect of the
bulk superconductor self-energy is introducing the $\Delta'\tilde{\tau}_{1}$
term defined by $\Delta'=\text{Tr(\ensuremath{\tilde{\tau}_{1}\hat{\Sigma}_{S}})}/2$, which is finite and proportional to $\alpha p_F/2S_F$ and $\Delta$.
The quasiclassical Green function $\hat{g}$ can be obtained from
Eq. \eqref{Eq:EilenbergerLeft} in the dirty limit as it is uniform and isotropic in the left part, $\boldsymbol{\nabla}\hat{g}=0$. Since the
bulk superconductor does not change any property of the 2D electron
layer except introducing the pairing correlation, the junction at
$x=0$ can be regarded as totally transparent. Therefore,
we can use continuous boundary condition at $x=0$, which is $\hat{g}(x=0^{+})=\hat{g}(x<0)$.
Then $f_{0}$ can be straightforwardly read out from $\hat{g}(0^{+})$. Substituting the expression of $f_0$ back into Eq.~\eqref{Eq:solution1} and \eqref{Eq:solution2}, we get the complete form of the proximity induced pairing correlations.

\section{Experimental Detection}

\begin{figure}[h!]
\centering
\includegraphics[width = 1\columnwidth]{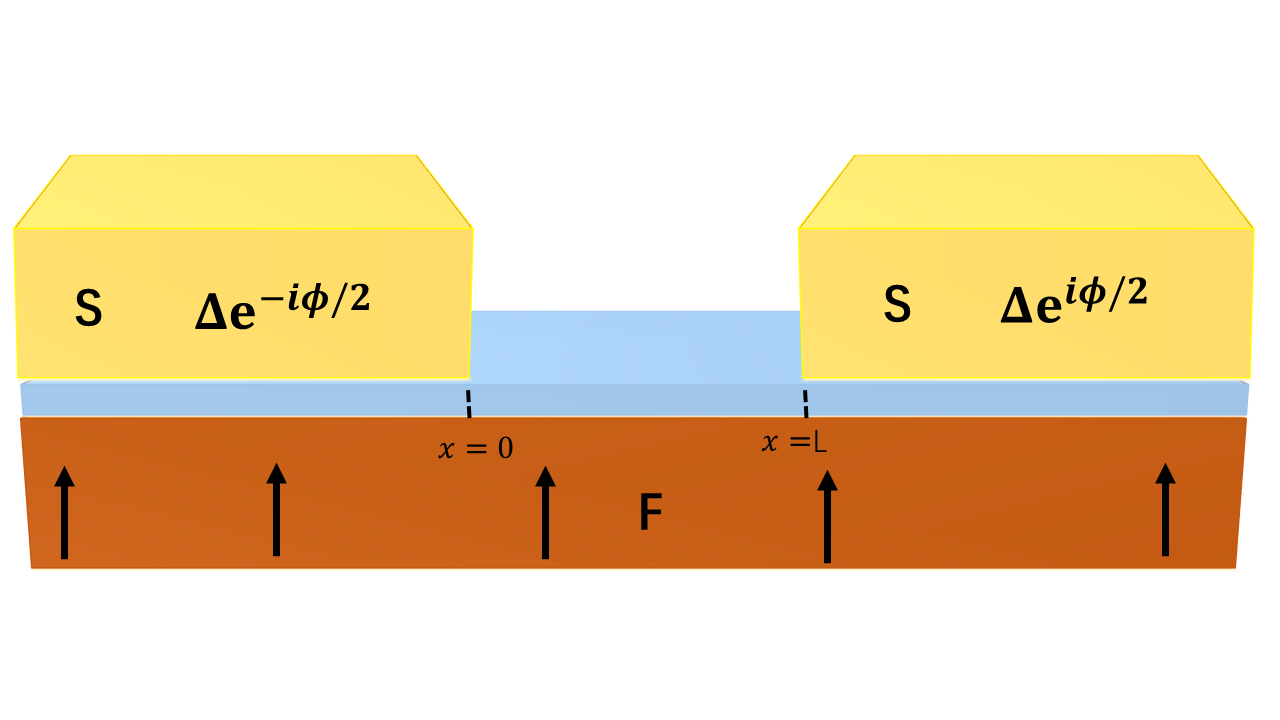}
\caption{The setup in which the supercurrent is measured. The two superconductors have the same pairing amplitude and a phase difference $\phi$.}\label{Fig:Is}
\end{figure}

The proximity effect can be detected experimentally by measuring the Josephson current through an S/F/S junction in a setup as shown in Fig. \ref{Fig:Is}. For simplicity, we assume that the two superconductors have the same pairing amplitude but with a phase difference. We first consider the case of $\alpha p_F\gg h_{z}$. In weak proximity limit, making use of Eq. \eqref{Eq:solution1} and \eqref{Eq:solution2} and matching the boundary conditions on the two sides $f(x=0)=f_{0}e^{-i\phi/2}$, $f(x=L)=f_{0}e^{i\phi/2}$, we get the induced pairing correlations  [\onlinecite{LRPEDirac}]

\begin{equation}
    f=\frac{f_{0}}{\sinh(\kappa L)}
    \begin{pmatrix}
    0 && A \\
    A^* && 0
    \end{pmatrix}
\end{equation}
with
\begin{eqnarray}
A&=&e^{\frac{2i\alpha p_{F}h_{y}(x-L)}{S_{F}v_{F}}+i\phi/2}\sinh(\kappa x)\nonumber\\
&-&e^{\frac{-2i\alpha p_{F}h_{y}x}{S_{F}v_{F}}-i\phi/2}\sinh[\kappa (x-L)],
\end{eqnarray}
where $L$ is the width of the junction. In the limit $\kappa L\gg 1$, the supercurrent density is calculated as

\begin{eqnarray}
I&=&-i\frac{\pi\sigma_{N}}{e}T\sum_{\omega_{n}>0}\text{Tr}\left[\tilde{\tau}_{3}\hat{g}\partial_{x}\hat{g}\right]\nonumber\\
&=&-\frac{\pi\sigma_{N} f_0^2}{e}\sin\left(\phi+\frac{2h_{y}\alpha p_{F}L}{S_{F}v_{F}}\right)T\sum_{\omega_{n}>0}\frac{\kappa}{\sinh(\kappa L)}\nonumber \\
&\approx&-\frac{2\pi\sigma_{N}f_0^2}{e}\sin\left(\phi+\frac{2h_{y}\alpha p_{F}L}{S_{F}v_{F}}\right)T\sum_{\omega_{n}>0}\kappa e^{-\kappa L},
\end{eqnarray}
where $\sigma_{N}=2e^2N_{0}D$ is the normal-state conductivity and the lower line is valid for $\kappa L \gg 1$. At low temperatures $T\approx 0$, the frequency summation can be converted to an integral
\begin{eqnarray}
I&=&-\frac{2\pi\sigma_{N}f_0^2}{e}\sin\left(\phi+\frac{2h_{y}\alpha p_{F}L}{S_{F}v_{F}}\right)\int d\omega\,\kappa e^{-\kappa L}\nonumber\\
&=&-\frac{8\pi D\sigma_{N}f_0^2}{eL^3}\sin\left(\phi+\frac{2h_{y}\alpha p_{F}L}{S_{F}v_{F}}\right)\nonumber\\
&&(CL^2+2\sqrt{C}L+2)e^{-\sqrt{C}L}.
\end{eqnarray}
Here, $C$ depends on the $x$ and $z$ components of the exchange field, $C=\frac{X}{4\tau D}+\frac{4h_{x}^2\alpha^2p_{F}^2}{S_{F}^2v_{F}^2}$. From the expression of the current, one can see that the critical supercurrent is suppressed by both $h_{z}$ and $h_{x}$. Another feature of the supercurrent is that the zero supercurrent state corresponds to a finite phase difference $\phi_{0}=-\frac{2h_{y}\alpha p_{F}L}{S_{F}v_{F}}$, which is due to breaking the inversion and mirror symmetries in the $x$ direction. At high temperature $T\gg 2DC,D/L^2$, the lowest frequency gives the dominating part to the current. Thus we have
\begin{eqnarray}
I\approx-\frac{2\pi\sigma_{N}f_0^2}{e}\sin\left(\phi+\frac{2h_{y}\alpha p_{F}L}{S_{F}v_{F}}\right)T\sqrt{\frac{\pi T}{2D}} e^{-\sqrt{\pi T/2D} L},
\end{eqnarray}
which does not depend on $h_{x}$ or $h_{z}$. In the case of $\alpha p_{F}\ll h_{z}$, we get the Josephson current using the same method

\begin{eqnarray}
I&=&-\frac{\pi D\sigma_{N}f_0^2}{2eL^3}\sin\left(\phi+\frac{2h_{y}\alpha p_{F}L}{S_{F}v_{F}}\right)\nonumber\\
&&(C'L^2+2\sqrt{C'}L+2)e^{-\sqrt{C'}L}
\end{eqnarray}
for $T\approx 0$, where $C'$ is given by $C'=\frac{Y}{\tau D}+\frac{4h_{x}^2\alpha^2p_{F}^2}{S_{F}^2v_{F}^2}$. At high temperatures, we have
\begin{eqnarray}
I\approx-\frac{2\pi\sigma_{N} f_0^2}{e}\sin\left(\phi+\frac{2h_{y}\alpha p_{F}L}{S_{F}v_{F}}\right)T\sqrt{\frac{2\pi T}{D}} e^{-\sqrt{2\pi T/2D} L}.
\end{eqnarray}
In the case of $\alpha p_{F}\approx h_{z}$, the Josephson current vanishes as there is almost no proximity effect.

\begin{figure}[h!]
\centering
\includegraphics[width = 1\columnwidth]{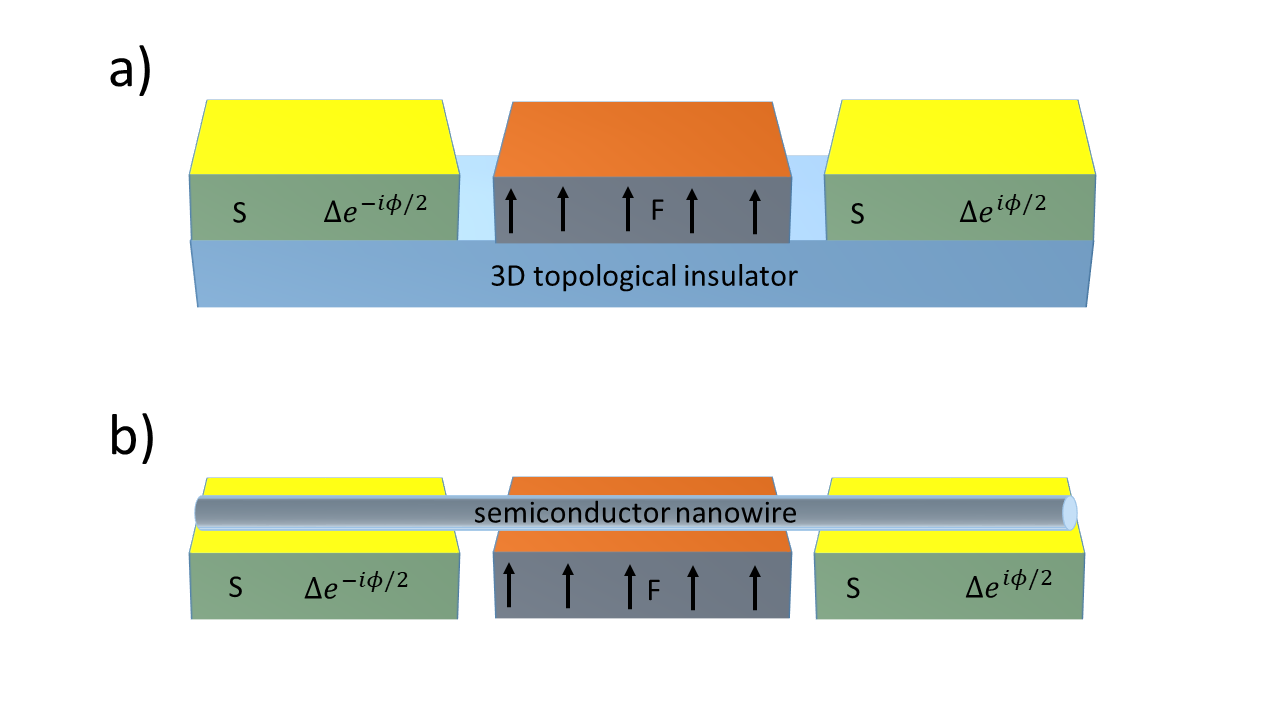}
\caption{Suggested experimental realizations: a) The case $\alpha p_F\ll h_z$ is realized when the bulk superconductors and a ferromagnet are put on the top of a 3D TI. A barrier is created between the ferromagnet and TI, such that the induced exchange field is small. (b) The case $\alpha p_F\ll h_z$ can be obtained with a semiconductor nanowire placed on  top of bulk superconductors and a ferromagnet, such that the induced exchange field is large compared with the Fermi energy of the nanowire.}\label{Fig:Ex}
\end{figure}

Let us briefly discuss the prospects of realizing our predictions experimentally in the presence of both spin-orbit coupling and exchange field, but in the two regimes where one is stronger than the other. In the case of $\alpha p_F\gg h_z$, we propose that one can use the surface of a doped 3D topological insulator (TI) such as $\text{Bi}_2\text{Se}_3$ [\onlinecite{TI1,TI2,TITSC}] as the electron layer with bulk superconductors and a ferromagnet on top of it [Fig.~\ref{Fig:Ex}(a)]. This system has a similar Fermi surface as that in Fig.~\ref{Fig:Spin}(b), so that the proximity effect is described by Eq.~\eqref{Eq:solution1}. The case $\alpha p_F\ll h_z$ can be realized by using [\onlinecite{Nw1,Nw2}] multi-channel semiconductor nanowires such as InAs as the electron layer put on top of superconductors and a ferromagnet [Fig.~\ref{Fig:Ex}(b)]. Although we do theoretical calculations in two dimensions, our results are valid for quasi-one dimensional systems as the proximity effect is homogeneous in the $y$ direction.

\section{CONCLUSION AND DISCUSSION}

\begin{figure}[h!]
\centering
\includegraphics[width = 1\columnwidth]{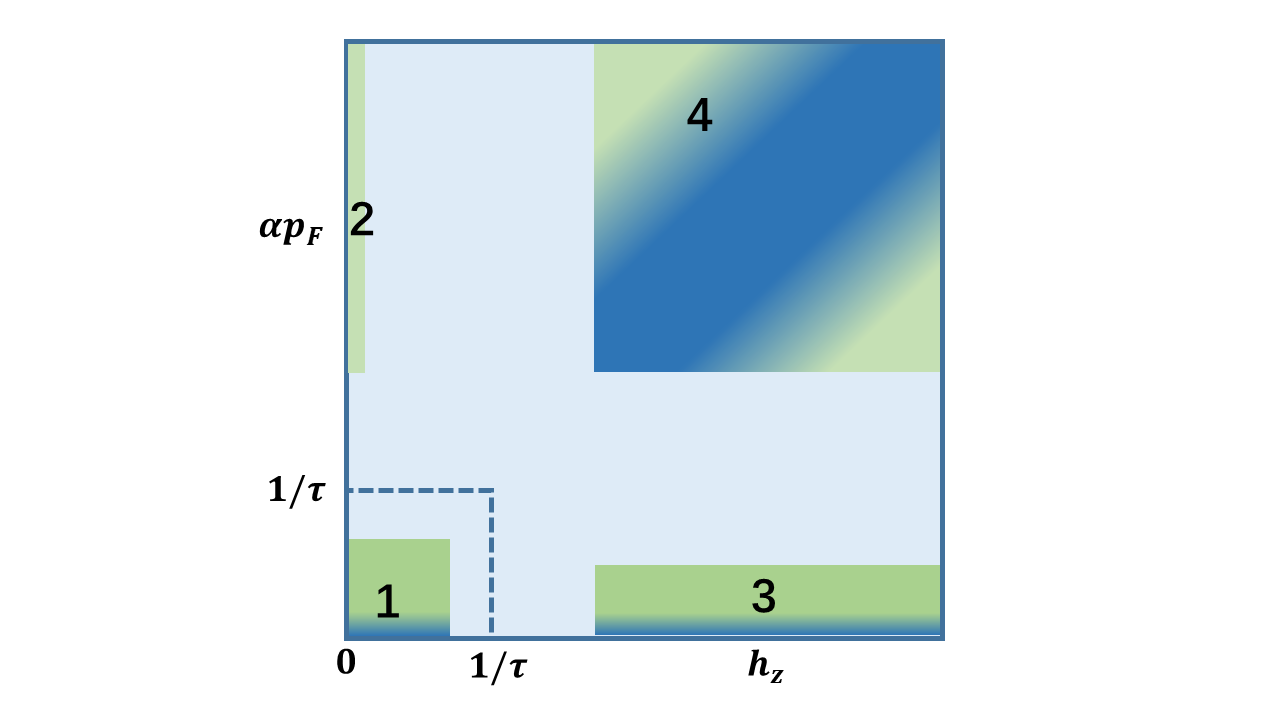}
\caption{Comparison of proximity effects in different parameter regimes. The blue (dark) areas correspond to the regimes in which there is no LRPE and the green (light) areas correspond to the regimes where there is LRPE. (1) $\alpha p_F,h_z\ll 1/\tau$ studied in Ref. [\onlinecite{LRSOC1,LRSOC2}], there is LRPE as long as $\alpha p_F$ is finite. (2) $\alpha p_F\gg 1/\tau$ and $h_z=0$ studied in Ref. [\onlinecite{LRPEDirac}], there is always LRPE. (3) $\alpha p_F\ll 1/\tau$ and $h_z\gg 1/\tau$ studied in Ref. [\onlinecite{Mihail}], there is LRPE as long as $\alpha p_F$ is finite. (4) $\alpha p_F,h_z\gg 1/\tau$ studied in the present work, LRPE exists in the $\alpha p_F\gg h_z$ and $\alpha p_F\ll h_z$ limits.}\label{Fig:conclusion}
\end{figure}

In conclusion, we consider the superconducting proximity effect in
a 2D electron layer with large SOC and exchange field and assume that
only one band is occupied. We derive a generalized quasiclassical
theory for this system by projecting the Green function onto the POB.
We show that the LRPE exists in $\alpha p_{F}\gg h_{z}$ and $h_{z}\gg\alpha p_{F}$
limits while when $\alpha p_{F}\approx h_{z}$ there is no LRPE. Our work fills the theoretical research gap of studying proximity effect in systems with both large SOC and exchange fields. Our work is compared with previous works on proximity effect in other parameter regimes in Fig.~\ref{Fig:conclusion}. Alhough
we study a specific model with large SOC and exchange field, our results
also apply to the surface of a three dimensional topological insulator in the presence of an
exchange field, and a quantum anomalous hall insulator because all these
models have a similar spin texture of the Fermi surface. Our method is
straightforward to generalize to any system with only
one large non-degenerate Fermi surface or systems with multiple Fermi
surfaces but very small inter-band scattering such as doped Weyl semimetals.

\begin{acknowledgments}
We thank M. Silaev and P. Virtanen for fruitful discussions. This work
is supported by the Academy of Finland project HYNEQ (Project no. 317118).
\end{acknowledgments}

\onecolumngrid
\appendix
\section{Calculation of $g_0$ in the case of $\alpha p_F\approx h_z$}

In this appendix, we calculate $g_0$ in the cases of $\alpha p_F\approx h_z$. In general $g_0$ can be written as
\begin{eqnarray}
g_{0}=\sum g_{0,i}(\phi)\tilde{\tau}_{i}.
\end{eqnarray}

In the diffusive limit, the pairing amplitude is independent of $\phi$, such that 
\begin{eqnarray}
g_{0,i}=|g_{0,i}|e^{if(\phi)},
\end{eqnarray}
where $f(\phi)$ is an arbitrary function. Using the condition $g_{0,i}(\phi)=g_{0,i}(\phi+2\pi)$, we find that $f(\phi)$ can only be $f(\phi)=iN\phi$ with $N\in Z$. Therefore, we have 

\begin{eqnarray}
g_{0,i}=\sum_{i}|g_{0,i}|e^{iN\phi}\tilde{\tau}_{i}.
\end{eqnarray}

The Pauli matrices $\tilde{\boldsymbol{\tau}}$ written in usual spin$\otimes$particle-hole basis are given by

\begin{eqnarray}
\tilde{\tau}_{1}&\hat{=}&\begin{pmatrix}\begin{array}{cccc}
0 & 0 & -\alpha p_{F} & (S_{F}-h_{z})e^{i\phi}\\
0 & 0 & (S_{F}+h_{z})e^{-i\phi} & -\alpha p_{F}\\
-\alpha p_{F} & (S_{F}+h_{z})e^{i\phi} & 0 & 0\\
(S_{F}-h_{z})e^{-i\phi} & -\alpha p_{F} & 0 & 0
\end{array}\end{pmatrix}/2S_{F}\nonumber\\
\tilde{\tau}_{2}&\hat{=}&\begin{pmatrix}\begin{array}{cccc}
0 & 0 & i\alpha p_{F} & -i(S_{F}-h_{z})e^{i\phi}\\
0 & 0 & -i(S_{F}+h_{z})e^{-i\phi} & i\alpha p_{F}\\
-i\alpha p_{F} & i(S_{F}+h_{z})e^{i\phi} & 0 & 0\\
i(S_{F}-h_{z})e^{-i\phi} & -i\alpha p_{F} & 0 & 0
\end{array}\end{pmatrix}/2S_{F}\nonumber\\
\tilde{\tau}_{3}&\hat{=}&\begin{pmatrix}\begin{array}{cccc}
S_{F}+h_{z} & -\alpha p_{F}e^{i\phi} & 0 & 0\\
-\alpha p_{F}e^{-i\phi} & S_{F}-h_{z} & 0 & 0\\
0 & 0 & -S_{F}+h_{z} & \alpha p_{F}e^{i\phi}\\
0 & 0 & \alpha p_{F}e^{-i\phi} & -S_{F}-h_{z}
\end{array}\end{pmatrix}/2S_{F}
\end{eqnarray}

It can be seen that for $|N|>1$, $\langle e^{iN\phi}\tilde{\tau}_{i}\rangle=0$. This means that $g_{0,i}$ only contains terms with $N=-1,0,1$. Checking all the possibilities of N, we find that the only solution for $g_{0}$ is $g_{0,3}=g_{s,3}$, $g_{0,1}=g_{0,2}=0$ where $g_{s,3}$ is a $\phi$ independent constant.

\section{Calculation of $\langle g_0\rangle$ in the cases of $\alpha p_F\gg h_z$ and $\alpha p_F\ll h_z$}

In this appendix, we calculate the angular average of $g_0$. First, we consider the case of $\alpha p_F\gg h_z$. According to Eq. (A4), we can get the angular average of $g_{0,i}\tilde{\tau}_{i}$ in the usual spin$\otimes$particle-hole basis,

\begin{eqnarray}
\langle g_{0,1}\tilde{\tau}_{1}\rangle&\hat{=}&g_{0,1}\begin{pmatrix}\begin{array}{cccc}
0 & 0 & -\alpha p_{F} & 0\\
0 & 0 & 0 & -\alpha p_{F}\\
-\alpha p_{F} & 0 & 0 & 0\\
0 & -\alpha p_{F} & 0 & 0
\end{array}\end{pmatrix}/2S_{F}\nonumber\\
\langle g_{0,2}\tilde{\tau}_{2}\rangle&\hat{=}&g_{0,2}\begin{pmatrix}\begin{array}{cccc}
0 & 0 & i\alpha p_{F} & 0\\
0 & 0 & 0 & i\alpha p_{F}\\
-i\alpha p_{F} & 0 & 0 & 0\\
0 & -i\alpha p_{F} & 0 & 0
\end{array}\end{pmatrix}/2S_{F}\nonumber \\
\langle g_{s,3}\tilde{\tau}_{3}\rangle&\hat{=}&g_{s,3}\begin{pmatrix}\begin{array}{cccc}
S_{F}-h_{z} & 0 & 0 & 0\\
0 & S_{F}+h_{z} & 0 & 0\\
0 & 0 & -S_{F}-h_{z} & 0\\
0 & 0 & 0 & -S_{F}+h_{z}
\end{array}\end{pmatrix}/2S_{F}.
\end{eqnarray}
Projecting $\langle g_{0,i}\tilde{\tau}_{i}\rangle$ back onto the $\psi_{-}$ basis, we obtain
\begin{eqnarray}
\langle g_{s,1}\tilde{\tau}_{1}\rangle&=&\Bigg(\frac{1}{2}-\frac{h_{z}^{2}}{2S_{F}^{2}}\Bigg)g_{s,1}\tilde{\tau}_{1}\nonumber \\ \langle g_{s,2}\tilde{\tau}_{2}\rangle&=&\Bigg(\frac{1}{2}-\frac{h_{z}^{2}}{2S_{F}^{2}}\Bigg)g_{s,2}\tilde{\tau}_{2} \nonumber \\
\langle g_{s,3}\tilde{\tau}_{3}\rangle&=&\Bigg(\frac{1}{2}+\frac{h_{z}^{2}}{2S_{F}^{2}}\Bigg)g_{s,3}\tilde{\tau}_{3}.
\end{eqnarray}
Therefore, we have
\begin{eqnarray}
\langle g_{s}\rangle=\frac{1}{2}g_{0}+\frac{X}{2}\tilde{\tau}_{3}g_{0}\tilde{\tau}_{3}
\end{eqnarray}
with $X=\frac{h_{z}^2}{S_{F}^2}$.

Next, we consider $\alpha p_F\ll h_z$. In this case, $\langle g_{s,i}\tilde{\tau}'_{i}\rangle$ written in the usual spin$\otimes$particle-hole basis is given by

\begin{eqnarray}
\langle g_{0,1}\tilde{\tau}'_{1}\rangle&\hat{=}&g_{0,1}\begin{pmatrix}\begin{array}{cccc}
0 & 0 & 0 & 0\\
0 & 0 & S_{F}+h_{z} & 0\\
0 & 0 & 0 & 0\\
S_{F}+h_{z} & 0 & 0 & 0
\end{array}\end{pmatrix}/2S_{F}\nonumber\\
\langle g_{0,2}\tilde{\tau}'_{2}\rangle&\hat{=}&g_{0,2}\begin{pmatrix}\begin{array}{cccc}
0 & 0 & 0 & 0\\
0 & 0 & -iS_{F}-ih_{z} & 0\\
0 & 0 & 0 & 0\\
iS_{F}+ih_{z} & 0 & 0 & 0
\end{array}\end{pmatrix}/2S_{F}\nonumber \\
\langle g_{s,3}\tilde{\tau}'_{3}\rangle&\hat{=}&g_{s,3}\begin{pmatrix}\begin{array}{cccc}
S_{F}-h_{z} & 0 & 0 & 0\\
0 & S_{F}+h_{z} & 0 & 0\\
0 & 0 & -S_{F}-h_{z} & 0\\
0 & 0 & 0 & -S_{F}+h_{z}
\end{array}\end{pmatrix}/2S_{F}
\end{eqnarray}
Projecting $\langle g_{0,i}\tilde{\tau}'_{i}\rangle$ back onto the $\psi_{-}^{'}$ basis, we obtain
\begin{eqnarray}
\langle g_{s,1}\tilde{\tau}'_{1}\rangle&=&\frac{(S_{F}+h_{z})^2}{4S_{F}^2}g_{s,1}\tilde{\tau}'_{1}\nonumber \\ \langle g_{s,2}\tilde{\tau}'_{2}\rangle&=&\frac{(S_{F}+h_{z})^2}{4S_{F}^2}g_{s,2}\tilde{\tau}'_{2} \nonumber \\
\langle g_{s,3}\tilde{\tau}'_{3}\rangle&=&\Bigg(\frac{1}{2}+\frac{h_{z}^{2}}{2S_{F}^{2}}\Bigg)g_{s,3}\tilde{\tau}'_{3}
\end{eqnarray}
Therefore, we have

\begin{eqnarray}
\langle g_{s}\rangle\approx g_{0}+\frac{Y}{2}\tilde{\tau}'_{3}g_{0}\tilde{\tau}'_{3}
\end{eqnarray}
with $Y=\frac{(S_{F}-h_{z})^2}{4S_{F}^2}$.

\end{document}